\documentclass[amssymb,prb,aps,showpacs]{revtex4}

\usepackage{epsfig}

\begin{document}
\title{Temporal dynamics of tunneling. Hydrodynamic approach.}
\author{G. Dekel, V. Fleurov\footnote{Email:fleurov@post.tau.ac.il.}}
\affiliation{Raymond and Beverly Sackler Faculty of Exact Sciences,
School of Physics and Astronomy,\\ Tel-Aviv University, Tel-Aviv
69978 Israel.}
\author{A. Soffer, C. Stucchio}
\affiliation{Department of Mathematics, Rutgers University, New
Brunswick, NJ 08903,USA}
\begin{abstract}
We use the hydrodynamic representation of the Gross -Pitaevskii/Nonlinear Schr\"odinger equation in order to analyze the dynamics of macroscopic tunneling process. We observe a tendency to a wave breaking and shock formation during the early stages of the tunneling process. A blip in the density distribution appears in the outskirts of the barrier and under proper conditions it may transform into a bright soliton. Our approach, based on the theory of shock formation in solutions of Burgers equation, allows us to find the parameters of the ejected blip (or soliton if formed) including the velocity of its propagation. The blip in the density is formed regardless of the value and sign of the nonlinearity parameter. However a soliton may be formed only if this parameter is negative (attraction) and large enough. A criterion is proposed. An ejection of a soliton is also observed numerically. We demonstrate, theoretically and numerically, controlled formation of soliton through tunneling. The mass of the ejected soliton is controlled by the initial state.
\end{abstract}

\maketitle

\section{Introduction}

The recent advances in the experiments on real Bose -- Einstein condensates(BEC) \cite{aemw95,dmaddkk95,bsth95}  and nonlinear optical waves\cite{fsec03,Agrawal} have generated a huge body of works on the theoretical side based on the Gross-Pitaevskii (GP)\cite{p61,g63} (see also Refs. \onlinecite{iasmsk98,ss99,l9800}) or Nonlinear Schr\"odinger (NLS) equation\cite{ss99,remark}
\begin{equation}\label{GPE}
i \hbar \frac{\partial}{\partial t} \Psi({\bf r}, t) = \left[ -
\frac{\hbar^2}{2 m} \nabla^2 + U({\bf r}) + \lambda |\Psi({\bf r},
t)|^2\right] \Psi({\bf r}, t).
\end{equation}

The dynamics of solutions of this equation are very complex and rich. The phenomena of coherence\cite{rp02}, macroscopic tunneling\cite{sspkpl04,chm05}, vortex formation,\cite{fpr92,wh99,mahhwc99,mcwd99,igrcgglpk01,a02} instabilities, focusing and blowup are all concepts related to the {\em nonlinear} nature of the systems. Dark solitons, or kink-wise states, i.e. states with dynamically stable propagating density minima, are expected in condensates with repulsive interactions. They have been predicted for one dimensional BEC \cite{morgan,reinhardt,jackson,muryshev} and may occur in higher dimensions as well. There have been several suggestions for techniques to engineer dark solitons in BEC \cite{dum} and these were indeed successfully created and observed \cite{burger1,denschlag}. The two groups used far off resonance laser beam pulse to shift the matter wave phase thus creating density minima.

As for bright solitons, the situation is more subtle, since in such systems instability of the gas is unavoidable above a critical particle number where the zero point kinetic energy does not suffice to balance the collapse mechanism. The latter renders the system to behaviors, which deviate significantly from the mean field validity range. The critical number has been calculated theoretically for $^7$Li to be $N_{cr}$ = 1400 \cite{dodd,dalfovo}, which is consistent with the experimental measurements. A condensate with a limited number of atoms, however, can be stabilized by confinement in a trap. In 1995, the first evidences for a BEC in $^7$Li atomic gas with attractive interactions were reported \cite{bradly}. Later that year, the stability of solitons created in condensates in a small harmonic trap, constrained to one dimensional motion, was predicted from numerical calculations \cite{ruprecht} For the 3d case, however, the solitons were predicted to be stable for modest ranges of nonlinearities. In 1998 an analytic solution was obtained for a BEC bright soliton creation in an asymmetric, cigar shaped trap. It was then shown that the solitons do not expand when confinement in one direction is lifted\cite{perez}. A major stepping stone was undoubtedly the production and observation by two experimental groups \cite{lkb,strecker1} of bright solitons in ultra cold $^7$Li gas, released from a one dimensional trap. Both groups reported propagation without dispersion over macroscopic distances. The latter group also observed a propagation of a soliton train.

The question of BEC bright soliton formation, stability and dynamics is far from being solved and the ongoing research is very active. Ways to stabilize two dimensional BEC solitons by using spatial modulation of the interaction strength \cite{montesinos} and by using rapid oscillations between repulsive and attractive
interactions\cite{saito} were suggested. The existence of vortex solitons in periodic potentials (optical lattices) was revealed in Refs. \onlinecite{kevrekidis,martika,alexander}. Exact solutions for the
dynamics of 1-D trapped BEC bright solitons with a time dependent interaction strength were found in Ref. \onlinecite{kanamoto}. Interference of tunneling BEC matter waves in an optical array was observed \cite{anderson} and existence of bounce solutions in macroscopic tunneling was investigated in Ref. \onlinecite{yasui}. Dynamics of an initially nonuniform bright soliton was studied in \cite{carr} as was friction and diffusion of bright solitons by the thermal cloud \cite{sinha}. Dynamics of a bright BEC soliton in an expulsive potential were investigated in \cite{salasnich}. Finally, pulsed macroscopic quantum tunneling of BEC are expected, which are induced by scattering of soliton on Guassian potential barrier \cite{spr01}.

Remarkably, many similar phenomena are observed in light propagation, when we have to turn to electromagnetic wave propagation and penetration into media with different refraction indices instead of matter wave dynamics. Atomic correlations now correspond to the coherence of laser light, while many-body (mean-field) interactions correspond to the Kerr nonlinearity. Examples of parallel dynamics
include soliton formation and modulation instability \cite{strecker1} in the focusing (attractive) case and dispersive shock waves in the defocusing (repulsive) case.\cite{dbsh01,wjf06} Multi-species condensates relate to multi-component, or vector, beams of light, while periodic potentials for both the atomic and photonic systems have been demonstrated using standing light waves.\cite{cbfmmtsi01,fsec03} Of course, there are also significant differences between the two systems, particularly when atomic excitations and quantum (vs. classical) statistics are involved. In these cases, too, it is useful to contrast optics with BEC in order to better understand the underlying dynamics of both.

Most of the theoretical analysis of the works mentioned above have so far been dealt with by a combination of numerical schemes (e.g. Ref. \onlinecite{sspkpl04,spr01}) and finite dimensional phenomenological models. Furthermore, all assume long time existence of the solitons in the gas. None has considered the problem of tracing the mechanisms responsible for the actual formation of the soliton in the course of tunneling. In other words, the problem of the short time dynamics of tunneling has so far not been addressed. The tendency to shock formation and creation of soliton at the early stages of the tunneling process is another new aspect of the theory presented in this paper. As we shall see below an interesting aspect of this process is a possibility to control such soliton parameters as mass, geometrical factors and velocity.

The hydrodynamic formulation for the Schr\"odinger equation was originally proposed in Ref. \onlinecite{m27}. A similar approach is also well-known in the linear and nonlinear optics (see, e.g. \cite{m75,s90}). Recently hydrodynamic formalism received much attention\cite{r05,flpz,carles,fs05,hacces06} as a useful tool for analyzing the GP/NLS equation. The time independent problem has been studied as far back as the early 1950's.\cite{a52} We apply this approach to one-dimensional systems. Its generalization to higher dimensions is straightforward although may require a special consideration of vortices.

We study nonlinear phenomena in macroscopic tunneling of a BEC gas or optical systems. Employing the hydrodynamic representation we analyze the time dependent GP/NLS equation (\ref{GPE}) and obtain the dynamics of a trapped droplet tunneling through a barrier both on the short and long time scales. We predict a splitting process, in which a blip in the density distribution is formed at short times
outside the confining potential. We find the conditions, under which it may evolve into an outgoing bright soliton. Our approach allows for an analytical calculation of its parameters including the velocity and energy. We also show a numerical evidence for the blip and soliton formation. This theory allows one to design a structure, in which we can fully control the parameters of the ejected soliton,
including its velocity and mass fraction split off of the initially trapped BEC. The latter observation also indicates a way to extract a stable BEC soliton out of a less stable one. These are feasible processes. Their experimental implementation may be carried out, e.g. by measuring light propagation in samples with spatially modulated refraction index\cite{Bar-ad,mls05}.

\section{General approach}
\label{procedure}

Eq. (\ref{GPE}) for a complex wave function $\Psi(x,t)$ is equivalently written as two equations for two real functions: the continuity equation
\begin{equation}\label{4}
\frac{\partial}{\partial t} \rho(x,t) + \frac{\partial}{\partial x
}[\rho(x,t)v(x,t)] = 0.
\end{equation}
for the particle density distribution $\rho(x,t)= |\psi(x,t)|^2$ and the Euler-type equation
\begin{widetext}
\begin{equation}\label{5}
\frac{\partial }{\partial} v(x,t) + v(x,t) \frac{\partial}{\partial x} v(x,t) = - \frac{1}{m}  \frac{\partial}{\partial x} V_{eff}(x,t).
\end{equation}
\end{widetext}
for the velocity field
$
v(x,t) = \frac{\partial}{\partial x} \phi,
$
where $\phi$ is the phase of wave function and
$$
V_{eff}(x,t) = V_{ext}(x) + V_{qu}(\{\rho(x,t)\}) + \lambda \rho(x,t)
$$
Here
\begin{equation}\label{QP}
V_{qu}(\{\rho(x,t)\}) = - \frac{\hbar^2}{2 m} \frac{1}{\sqrt{\rho(x,t)}} \frac{\partial^2}{\partial x^2} \sqrt{\rho(x,t)}
\end{equation}
is quantum potential (called also quantum pressure) (QP). The effective potential $V_{eff}(x,t)$ is a functional of the density distribution $\rho(x,t)$ and may vary in time so that generally the two equation should be analyzed together.

Sometimes we have information about the density distribution and may solve Eq. (\ref{5}) separately. The diffusion term
$$
\frac{\zeta}{2m} \frac{\partial^2}{\partial x^2} v(x,t)
$$
in the right hand side of Eq. (\ref{5}) is introduced as a singular perturbation\cite{s80} which ensures its dissipative regularization,
\begin{widetext}
\begin{equation}\label{5-d}
\frac{\partial }{\partial} v(x,t) + v(x,t) \frac{\partial}{\partial x} v(x,t) = \frac{\zeta}{2m} \frac{\partial^2}{\partial x^2} v(x,t) - \frac{1}{m}  \frac{\partial}{\partial x} V_{eff}(x,t).
\end{equation}
\end{widetext}
We expect a tendency to formation of a shock wave in the velocity and density distributions. It is well know that the Burgers equation with the diffusion term has step-like solutions, which survive even if the 'diffusion' coefficient $\zeta$ tends to zero. However this type of a solution does not appear if $\zeta = 0$ from the very beginning (see e.g., Eqs. (4.1) and (4.2) in Ref. \onlinecite{hacces06}). Therefore we first analyze Eq. (\ref{5-d}) for finite values of the coefficient $\zeta$  and then take the limit $\zeta \to 0$ in the final results.

We may now apply the Cole-Hopf transformation
\begin{equation}\label{CH}
v(x,t) = - \displaystyle\frac{\zeta}{m} \frac{\varphi_x(x,t)}{\varphi(x,t)}
\end{equation}
so that the new function $\varphi(x,t)$ satisfies the linear diffusion equation with a source
\begin{equation}\label{diff}
\varphi_t(x,t) = \frac{\zeta}{2m} \varphi_{xx}(x,t) +
\frac{1}{\zeta} V_{eff}(x,t) \varphi(x,t).
\end{equation}
The Green function of Eq. (\ref{diff}) can be represented by the Wiener path integral
\begin{equation}\label{Green1}
G(x,x_0,t,0)= \int_{x_0,0}^{x,t} D \left[x(\tau)\right]
e^{-\displaystyle \frac{1}{\zeta} S([x(\tau)];t,0)}
\end{equation}
where
$$
S([x(\tau)];t,0) = \int_{t_0}^t \left[ \frac{m}{2\zeta}\left(\frac{dx(\tau)}{d\tau} \right)^2d \tau - V_{eff}(x(\tau),\tau)\right]d\tau
$$
has the form of an action for a particle with the mass $m$ moving along a path in the potential $V_{eff}$. Distinguishing as usually the contribution of the classical path $x_c(\tau)$ we get the saddle point approximation for the Green function,
\begin{equation}
G(x, x_0;t,0)= F(t) e^{- \displaystyle \frac{1}{\zeta}
S([x_c(\tau)];t,0)}.
\end{equation}
This approximation leads in fact to exact results in the limit $\zeta \to 0$. In particular the pre-exponential factor $F(t)$ plays no role in this limit. We look for the solution $\varphi(x,t)$ of Eq. (\ref{diff}), which at $t = 0$ has the form
$$
\varphi_0(x) = e ^{- \displaystyle \frac{1}{\zeta} S_0(x)}
$$
Returning to the Cole-Hopf transformation (\ref{CH}) we understand
that the  function $S_0(x)$ must be chosen to satisfy the condition
$$
v_0(x) = \frac{1}{m} \frac{d S_0(x)}{dx}
$$
where $v_0(x)$ is the initial velocity field at $t = 0$ of the
original physical problem.

The solution of Eq. (\ref{diff}) with this initial condition then reads
\begin{widetext}
\begin{equation}\label{solution1}
\varphi(x,t)=\int_{-\infty}^{\infty} d x' G(x,x',t) \varphi_0(x') = F(t) \int_{-\infty}^\infty dx' e^{-\displaystyle \frac{1}{\xi} \left(S_0(x') + S(x,x',t)\right)}
\end{equation}
\end{widetext}
The integration in Eq. (\ref{solution1}) is carried out around the saddle point defined by
\begin{equation}\label{initial}
\frac{\partial S(x, x',t)}{\partial x'} = - m v_0(x')
\end{equation}
so that
\begin{equation}\label{solution2}
\varphi(x,t) \propto e^{-\displaystyle \frac{1}{\xi}  \left(S_0(\bar
x(x,t)) + S(x,\bar x(x,t),t)\right)}
\end{equation}
where $\bar x(x,t)$ is obtained by solving Eq. (\ref{initial}) with
respect to $x'$ for given $x$ and $t$. Substituting solution (\ref{solution2}) into Eq. (\ref{CH}), using the initial condition (\ref{initial}) and taking the limit $\zeta \to 0$ one gets the velocity field in the form
$$
v(x,t) = \frac{\partial S(x,\bar x(x,t),t)}{\partial x}
$$
where the last equality follows from Eq. (\ref{initial}).

This result has a simple interpretation. As mentioned above the function $S(x,\bar x,t)$ is the mechanical action of a particle (to be called 'tracer' below) with the mass $m$ moving from the point $\bar x$ to $x$ during the time $t$ in the potential $V_{eff}$. $v_0(\bar x)$ is its initial velocity. In other words we have to solve the following equivalent problem. At a time $t$ an observer measures the velocity field in an 'effective fluid' flowing in the external potential $V_{eff}(x,t)$ under the condition that the initial velocity field is $v_0(x)$. The possible compressibility of the fluid is accounted for by the dependence of the quantum pressure and $V_{eff}(x)$ on the density distribution. If the measurement is
carried out in the point $x$ the observer sees the tracer, which has started at the time 0 from a point $\bar x$ with the velocity $v_0(\bar x)$. The question is what the tracer's velocity measured by the observer is. Solving this problem we obtain the velocity distribution of the fluid flowing in the effective potential (\ref{5b}), which according to the above procedure coincides with the velocity field of the actual quantum fluid. Below we analyze this problem for a model choice of the effective potential $V_{eff}$ and obtain the velocity field as a function of time $t$.

\section{Model}

\subsection{Adiabatic approximation}

\begin{figure}[h]
\vspace{0.5cm}
\includegraphics[width=6cm,angle=0]{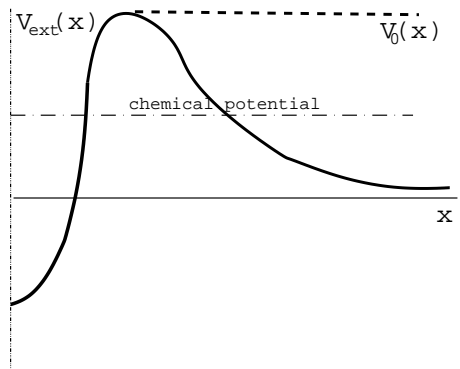}
\caption{The right half view of the trap potential $V_{ext}(x)$ keeping the droplet. The dashed line shows the auxiliary potential $V_0(x)$ which differs from $V_{ext}(x)$ only to the right from the top of the potential barrier. The dashed-dotted line corresponds to the chemical potential $\mu$ for a given number $N$ of particles.} \label{f.1}
\end{figure}

We consider here tunneling escape from the one-dimensional potential trap, Fig. \ref{f.1}.
Tunneling of a particle with the energy $\mu$ from the trap is characterized by the time of tunneling escape
$$
\tau_{tun} \sim \nu^{-1} e^\lambda
$$
where $\nu$ is the frequency of oscillations  within the trap, and
$$
\lambda = \frac{1}{\hbar}\int_a^b dx \sqrt{2m(V_{ext}(x) - \mu)}
$$
is the tunneling integral of the classically forbidden underbarrier
region $(a,b)$. There is also another important characteristic of tunneling which is the time $\tau_{tr}$ needed to traverse the under-barrier region\cite{hs89}. This duality in tunneling characterization was first introduced in Ref. \onlinecite{maccol} and has been dealt with in many works since\cite{hs89,pollak,balcou,bl82}. The traversal time is a matter of intensive research and attracts attention both from fundamental and technological reasons. Among many possible definitions of the traversal time the 'semiclassical' traversal time
\begin{equation}\label{traversal}
\tau_{tr} = \sqrt{\frac{m}{2}}\int_{a}^{b}dx \frac{1}{\sqrt{V_{ext}(x) -
E}}
\end{equation}
proposed in Ref. \onlinecite{bl82} is most popular and will be used below. It is obtained by modulating the potential barrier with a small cosine perturbation and finding the traversal time as the crossover between high and low frequency behaviors.

In what follows we embrace the definition (\ref{traversal}) for the traversal time and rely on the additional important observation that the density field varies on the time scale $\tau_{tun}$, whereas the velocity field varies on the timescale $\tau_{tr}$\cite{fs05}. The inequality $\tau_{tr} \ll \tau_{tun}$, valid for typical barriers, assumes dynamics of the velocity field at virtually constant
density. We may therefore use this assumption in order to apply the adiabatic approximation in the first iteration. The latter will end, by definition, when changes in the velocity field will become significant. The second step will be to calculate the time dependence of the density, $\rho(x,t)$, for the time dependent velocity field found in the first iteration.

We consider now a droplet trapped in the potential $V_{ext}(x)$, from which it escapes due to tunneling through the potential barrier. In order to define the initial density distribution $\rho_0(x)$ we introduce an auxiliary confining potential $V_{0}(x)$, from which tunneling is impossible and a stationary state is formed. This potential coincides with $V_{ext}(x)$ for small $x$ up to the top of the potential barrier but differs for larger $x$ (dashed line in fig. \ref{f.1}). The density distribution $\rho_0(x;N)$ of the stationary state in the auxiliary potential $V_{0}(x)$ is obtained from Eq. (\ref{5}) at $v(x,t)=0$, which becomes
\begin{equation}\label{6}
\left[ V_0(x) -  \frac{\hbar^2}{2 m} \frac{\partial^2}{\partial x^2} + \lambda \rho_0(x;N) \right]\sqrt{\rho_0(x;N)} = \mu(N) \sqrt{\rho_0(x;N)}.
\end{equation}
Eq. (\ref{6}) is in fact the stationary GP/NLS equation with the potential $V_0(x)$ and determines therefore the initial density distribution $\rho_0(x;N)$. Here $\mu(N)$ is the chemical potential. The calculations is carried out for a given total number $N$ of the particles in the trap.

Using Eq. (\ref{6}) for $\rho_0(x;N)$ we get that the effective potential in Eq. (\ref{5-d}) becomes for a given number $N(t)$ of particles becomes
\begin{widetext}
\begin{equation}\label{5b}
V_{eff}(x) = V_{ext}(x) - V_0(x) + \mu + V_{qu}(\{\rho(x,t)\})  -
V_{qu}(\{\rho_0(x;N(t))\}) + \lambda [\rho(x,t) - \rho_0(x;N(t))].
\end{equation}
\end{widetext}
The number of particles $N(t)$ in the trap may vary slowly with time. Generally the effective potential follows the variation of density distribution. However, the adiabatic approximation implies that the density field varies much slower than the velocity field. It ensures also that the QP follows also the change of the total number of particles in the trap in the course of time, which is reflected by a slow variation of the chemical potential $\mu$ with slowly changing number of particles $N(t)$.\cite{fs05} Hence we may, to within a good precision, assume that $\rho(x,t) = \rho_0(x;N(t))$ in Eq. (\ref{5b}), meaning that
\begin{equation}\label{5c}
V_{eff}(x) = V_{ext}(x) - V_0(x) - U_0.
\end{equation}
where $U_0$ is the asymptotic value of the difference $V_{ext}(x) - V_0(x)$ at large $x$. The chemical potential may be dropped under the derivative over the coordinate $x$ in Eq. (\ref{5-d}) and we may introduce for the sake of convenience the constant $- U_0$ so that $V_{eff}(x) \to 0$ at $x \to \infty$. Therefore the effective potential is the shifted by $U_0$ time independent difference of the actual potential forming the trap and the auxiliary potential used to form the initial density distribution.

Then it is convenient to chose the effective potential (\ref{5c}) in the form
\begin{equation}\label{effpot}
V_{eff}(x) = \frac{U_0}{\cosh^2 \alpha x}
\end{equation}
Since we start from the stationary state $\rho_0(x;N)$ as the initial density distribution, we also assume the zero initial velocity field, $v_0(x) = 0$. The effective potential (\ref{effpot}) is time-independent so that we deal with an 'adiabatically' incompressible flow. The compressibility, i.e. rapid variations of $\rho(x,t)$ and, hence, $V_{eff}$ with time can be accounted for in higher iterations.

According to the analysis in Section \ref{procedure} we have to calculate the velocity field $v(x,t)$ of the equivalent fluid at a given point $x$ and time $t$ with the initial conditions described above. For this sake we consider a fluid tracer which starts moving at a point $\bar x$ at the initial time $t=0$ with the initial velocity $v(x,0) = 0$. The initial energy of the tracer
\begin{equation}\label{conserv-1}
\bar \varepsilon = \frac{U_0}{\cosh^2 \alpha \bar x}.
\end{equation}
is conserved in the Lagrange coordinates, when we follow the tracer along the path of its motion. Hence at the point of observation $x$ (which is a static Euler coordinate) and at the time of observation $t$ it becomes
\begin{equation}\label{conserv-2}
\bar \varepsilon = \varepsilon(x,t) \equiv \frac{U_0}{\cosh^2 \alpha x} + \frac{m v^2(x,t)}{2}.
\end{equation}
Eqs. (\ref{conserv-1}) and (\ref{conserv-2}) allow us to connect the starting point $\bar x$ of the tracer motion with the velocity field measured at the time $t$ at the point $x$.

The tracer, which has started from the point $\bar x$ with the energy $\varepsilon$, reaches the point of observation $x$ after the time
\begin{equation}\label{time}
t = \int_{\bar x}^{x} \frac{1}{\sqrt{\displaystyle \frac{2}{m} \left(\varepsilon -  \frac{U_0}{\cosh^2 \alpha x}\right)}}\ dx.
\end{equation}
This integral together with the initial condition (\ref{conserv-1}) implicitly defines the energy $\varepsilon(x,t)$ of the tracer observed in the point $x$ at the time $t$. Calculating the integral in Eq. (\ref{time}) we arrive at equation
\begin{equation}\label{w}
F(w;\xi,\tau) \equiv 2 - w + 2 \sqrt{1 - w} - w \exp\left\{\frac{\tau}{\sqrt{w \sinh^2\xi
+ 1}}\right\} = 0.
\end{equation}
where
$$
w = \frac{U_0 - \varepsilon}{\varepsilon \sinh^{2}\xi}
$$
with $0\leq w \leq 1$. Here we use the dimensionless  time, $\tau = t \sqrt{8U_0 \alpha^2/m}$, and space, $\xi = \alpha x$, coordinates. It is worth noting that the time scale $\sqrt{m/2U_0\alpha^2}$ is of the order of the traversal tunneling time (\ref{traversal}) and appears as a natural scale for the time variation of the velocity field in the course of tunneling.

\subsection{Velocity field}
\label{velocity}

Eq. (\ref{w}) is solved with respect to $w$, so that we get the quantity $\varepsilon(\xi,\tau)$ at given $\xi$ and $\tau$. Then we obtain the velocity field
$$
v(\xi,\tau) = \sqrt{\frac{2}{m}\left [\varepsilon(\xi,\tau) - \frac{U_0}{\cosh^2 \xi} \right]}.
$$
Eq. (\ref{w}) is nonlinear and may have more than one solution. One can find the critical time $\tau_c =5.55$ and position $\xi_c = 2.005$ from the condition that the function $F(w;\xi,\tau)$ in Eq. (\ref{w}) zeros simultaneously with its first and second derivatives. Then for $\tau < \tau_c$, Eq. (\ref{w}) has only one solution at each value of the coordinate $\xi$. At longer times $\tau > \tau_c$, there is a finite range of $\xi$ values at $\xi > \xi_c$, where Eq. (\ref{w}) has three solutions.

The appearance of three solutions corresponds to breakdown of the wave (see, e.g. discussion in Ref. \onlinecite{w74}) and to formation of a shock wave. In the critical region the procedure as outlined in Section \ref{procedure} should be amended. It means in fact that Eq. (\ref{initial}) has several solutions and, hence, the integrand in Eq. (\ref{solution1}) has several saddle points. The saddle point, at which the action $S(x,x',t)$ is the smallest will determine the actual
velocity field in the limit $\zeta \to 0$.

\begin{figure}[h]
\includegraphics[width=6cm,angle=-0]{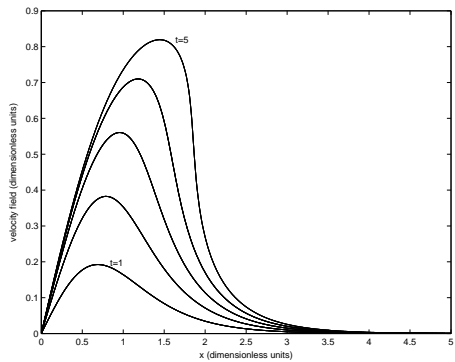}
\caption{Velocity profiles (in the $\sqrt{2U_0/m}$ units),
$\widetilde v(\xi)$, for increasing times, $\tau = 1,2,3,4,5$.}
\label{graphvelo-1}
\end{figure}

\begin{figure}[h]
\includegraphics[width=6cm,angle=-0]{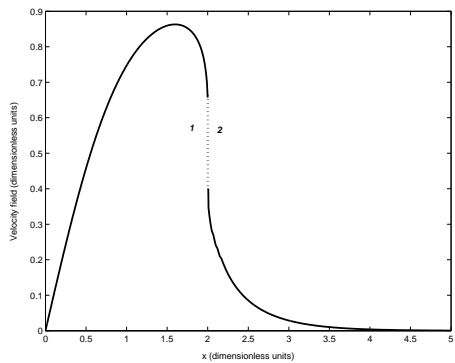}
\caption{Velocity profile (in the $\sqrt{2U_0/m}$ units) at $\tau=
\tau_c$ where breakdown of the wave occur.} \label{graphvelo-2}
\end{figure}

The velocity field at $\tau \leq \tau_c$ is shown in Fig. \ref{graphvelo-1}. A blip is formed on the outskirts of the barrier, which then approaches the wave breakdown at $\tau = \tau_c$ Fig. \ref{graphvelo-2}. This type of behavior is characteristic of the shock wave formation. We do not follow the further evolution of the velocity field since we have stretched the adiabatic approximation to its limit. In fact we should consider this result as the first iteration and calculate the evolution of the density field by means of the continuity equation (\ref{4}), which may introduce corrections to the effective potential (\ref{effpot}) in the region where the shock wave is going to be formed.

Although the fully developed shock with a sharp step in the density distribution would be an artifact of the approximation of incompressible flow used to calculate the velocity field, its analysis allows for an estimate of the speed of blip propagation. Solving simultaneously equations
$$
F(w;\xi,\tau) = 0, \ \ \ \mbox{and}\ \ \ \frac{\partial F(w;\xi,\tau)}{\partial\xi} = 0
$$
at a given time $\tau > \tau_c $ we find two values of the coordinate $\xi$  between which the shock occurs. Using the fact that $w$ is small in this region, we obtain the upper limit of the blip velocity
$$
v_b = \sqrt{\frac{2U_0}{m}}.
$$
It is worth emphasizing that $v_b$ is of the order of the velocity, with which the tunneling particles traverse the classically forbidden barrier region.

\subsection{Density field}
\label{density}

\begin{figure}
\includegraphics[width=7cm,angle=-0]{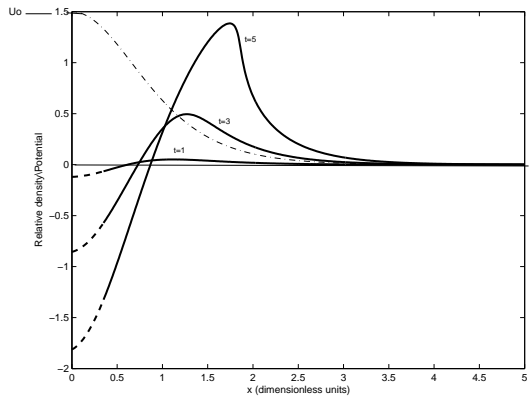}
\caption{The blip in the density distribution at $\tau=1,\ 3,\ 5$
for  $\beta = 0.25 \alpha$. The lines become dashed inside the trap
where the calculation errors may be large. The shape of the
effective potential (out of scale) is shown by the dash-dotted line.
} \label{blipwithtimepot}
\end{figure}

The development of a blip in the velocity field results in a local increase of the density at $\tau < \tau_c$ (see Fig. \ref{blipwithtimepot}). The variation of $\rho(x,t)$ near the blip is found from the continuity equation (\ref{4}) assuming similarly to Ref. \onlinecite{fs05} that in the region where the blip is formed (i.e. outside the trap) the initial density distribution is
$$
\rho_0(x)= \tilde\rho e^{ - \beta x}
$$
where $\beta = 2 \sqrt{2m(U_0 - \mu)}$ and $\tilde \rho$ is a constant.  We solve equation
\begin{equation}\label{rho1}
\frac{\partial \rho(x,t)}{\partial t} = - \frac{\partial}{\partial x}\left[\rho_0(x,t) v(x,t)\right].
\end{equation}
where $v(x,t)$ is the time dependent velocity field calculated in Subsection \ref{velocity}. The relative variation of the density
\begin{equation}\label{densvar}
\rho_{rel}(\xi, \tau) \equiv \frac{\rho(\xi,\tau) - \rho_0(\xi)}{\rho_0(\xi)} =
\int_0^{\tau} d\tau' \frac{1}{2} \left[ \frac{\beta}{\alpha} v(\xi,\tau') - \frac{\partial v(\xi,\tau')}{\partial \xi}\right]
\end{equation}
is obtained by integrating Eq. (\ref{rho1}) over time.

The calculated distribution $\rho_{rel}(\xi,\tau)$ of the relative density  variation is shown in Fig. \ref{blipwithtimepot} at several times $\tau < \tau_c$.
The blip in the density propagates with the velocity $v_b$ leaving a depleted region behind. Although the depletion is a real effect the approximate procedure leads to a too strong depletion inside the trap as shown by the dashed lines meaning that we cross the applicability limits of the procedure.

We clearly see a tendency to the wave breakdown and to formation of a shock at times approaching the critical time $\tau_c$. This however doesn't happen and the wave doesn't break down and form a shock in the density field $\rho(x)$, since such a development is inhibited by the quantum potential (\ref{QP}). The latter contains the second derivative of the density which may become large in the area of the
blip formation even though the absolute value of $\rho_{rel}(\xi,\tau)$ may be still small. The corresponding correction to the effective potential (\ref{5b}) prevents the sharp shock formation.

\subsubsection{Formation of a soliton}

At $\tau = 3$ the blip moves in a weak potential in the outskirts of the trap, and we deal with a new problem of a packet propagating with a the velocity $\sim v_b$, which may or may not transform into a soliton. The direct comparison shows that the shape of the blip is close to that of the soliton. The latter is formed if the blip energy $E_{blip}$ in its center of mass coordinate system is negative (e.g., Ref. \onlinecite{sw05} and references therein). This condition may be fulfilled only if the interaction parameter $\tilde \lambda = 2m\lambda/\hbar^2$ is negative (attractive interaction). Calculating $E_{blip}$ by means of the Gross-Pitaevskii surface energy functional we get the inequality
\begin{equation}\label{blipf}
|\tilde{\lambda}|\tilde\rho > (12.95\alpha^2 + 0.123 \frac{m
U_0}{2\hbar^2}),
\end{equation}
i.e. for each value of $U_0$ and $\alpha$ there is a lower limit for the interaction strength above which a soliton may be formed. Its width is about twice the width of the trap and it contains about 10\% of the initial packet. For the typical parameters of the currently available systems, $m = 7$ A.M.U = $11.69\cdot10^{-27}$kg (for Li atoms), $U_0=10^{-33}$J, $\alpha=10^4$ m$^{-1} $, and $\tilde\rho = 10^{16}$ m$^{-3}$ we get $|\tilde{\lambda}| > 2.38\cdot10^{-7} m$. Finally, a soliton is formed if the interaction coefficient satisfies the condition $|\lambda| > 0.73 \cdot10^{-49} J \cdot m^3$, which is typically fulfilled and has been measured experimentally (see, e.g. \cite{lkb}).

It is emphasized that in the case of a weaker negative (attractive), zero or even positive (repulsive) interaction parameter $\lambda$ the soliton is not formed, however the soliton-like blip in the density will be always formed and propagate far away from the trap (many hundreds of barrier widths) before being dispersed.

\subsection{Numerical solution}

\begin{figure}
\includegraphics[width=5.7cm,angle=-90]{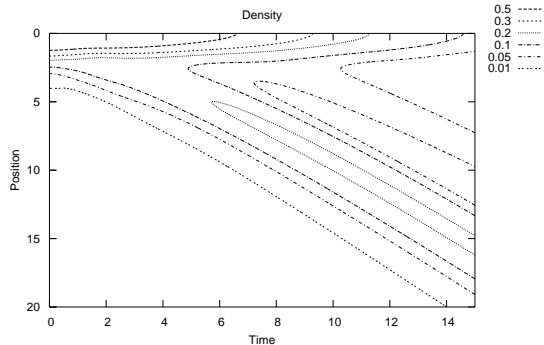}
\caption{Contour plot of the density distribution evolving with time. The figure shows formation of a blip outside the barrier at short times.}
\label{f.3}
\end{figure}

\begin{figure}
\includegraphics[width=5.7cm,angle=-90]{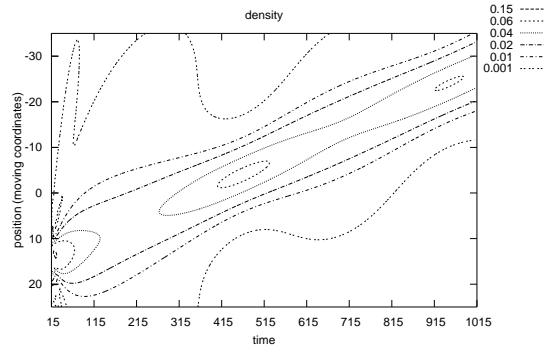}
\caption{Contour plot of the density distribution evolving  with
time. The figure shows propagation of the blip and its gradual
conversion into a soliton at long times. The lower graph uses the
frame, which moves with the velocity $v = 1.047 $ and follows the
blip.} \label{f.4}
\end{figure}

We may observe as soliton is formed and ejected in the course of tunneling also in the direct numerical solution of the GP/NLS Eq. (\ref{GPE}) carried out using the program "Kitty"\cite{sc06}. Varying the trap potential in a wide range of its parameters we were always able to observe formation of the blip. Here we present an example of the computation carried out for the GP/NLS hamiltonian
$$
H = - \frac{1}{2}\nabla^2 + V(x) - \frac{1}{2} |\Psi(x,t)|^2
$$
where
$$
V(x) = \frac{9}{8} \left(1 + \frac{x^4}{25}\right) \exp(-\frac{x^4}{35})
$$
with the initial wave function
$$
\Psi(x,0) = \frac{3\sqrt{2}}{4} \cosh \frac{3x}{4}.
$$
Formation of a blip with a negative energy outside the potential barrier at short times is shown in Fig. \ref{f.3} whereas Fig. \ref{f.4} shows its propagation at longer times. One can clearly see two parallel lines showing the core of the blip, which slowly oscillates approaching the soliton shape. It is worth emphasizing that the characteristics of the soliton, obtained numerically, are quite close to those obtained in our above analysis, including the time and location of the blip formation and the velocity of its propagation.

\section{Conclusions}

Using the fluid dynamics paradigm, we have analyzed the GP/NLS equation and demonstrated for the first time the phenomenon of a shock formation in the course of tunneling short time dynamics. It results in the formation and ejection of a blip in the density field of the outgoing mass. When the suitable conditions are fulfilled the latter may later on transform into a bright soliton. We propose a technique which allows for the analysis of the short time dynamics of tunneling
processes which provides us with a new insight to the fundamental problem of macroscopic tunneling.

We believe that the conclusion of this paper can be straightforwardly verified experimentally both in BEC tunneling or in nonlinear optics measurements. By engineering a trap with the proper parameters we control the tunneling process and formation of bright solitons. The blip splits from a bigger and narrower and thus less stable trapped packet, therefore a general approach of ensuring stability by the BEC tunneling is presented.

Although the model considered here assumes that a BEC droplet is bound in a trap, the blip formation in the course of tunneling is the property of the barrier and is observed for other initial states as well, e.g., numerics show a similar effect in 2d - systems.

The model we analyzed can be implemented experimentally both for BEC and optical soliton devices. The latter requires a quite simple setup of an optical fiber, with a spatially modulated refraction index, constituting the trap and its outings. As the light propagates, a blip should be detected near the trap outings, which escapes towards the sides of the fiber, and develops into a soliton. By controlling the initial state, i.e. the initial mass of the BEC droplet or light intensity, we may obtain a 'soliton gun' for prescribed mass and velocity of the ejected solitons. Our approach applies to other nonlinear soliton dynamical effects, e.g. soliton slicing\cite{hmz01} (by a potential bump) and generally soliton
interacting with a potential. Another important example  may be the dynamics of nonlinear models of fission. (see, e.g. \cite{nn04}).

\begin{acknowledgments}

The authors are indebted to S. Flach, S. Bar-Ad, J. Brand, and J.
Fleischer for useful comments. A.S. and G.D. are partially supported
by NSF. V.F. and G.D. are supported by Israeli Science Foundation,
grant No. 0900017. V.F. was partially supported by NSF grant
No. DMR-0442066 during his stay in Rutgers.

\end{acknowledgments}


\begin{thebibliography}{99}
\bibitem{remark} Eq. (\ref{GPE}) is written as an equation for a
quantum liquid. Changing notations the same equation describes light
propogation in a medium with the Kerr nonlinearity.
\end{thebibliography}
\end{document}